\documentclass[amsmath,11pt]{article}

\usepackage{amssymb,amssymb,epsfig,bm}
\usepackage{caption}
\usepackage{subcaption}

\date{\empty}

\begin{document}

\title{\bf Peculiar velocities in Friedmann universes with nonzero spatial curvature}
\author{Eleftheria P. Miliou${}^1$ and Christos G. Tsagas${}^{1,2}$\\ {\small ${}^1$Section of Astrophysics, Astronomy and Mechanics, Department of Physics}\\ {\small Aristotle
University of Thessaloniki, Thessaloniki 54124, Greece}\\ {\small ${}^2$Clare Hall, University of Cambridge, Herschel Road, Cambridge CB3 9AL, UK}}

\maketitle

\begin{abstract}
We extend the earlier linear studies of cosmological peculiar velocities to Friedmann universes with nonzero spatial curvature. In the process, we also compare our results with those obtained in cosmologies with Euclidean spatial sections. Employing relativistic cosmological perturbation theory, we first provide the differential formulae governing the evolution of peculiar velocities on all Friedmann backgrounds. The technical complexities of the curved models, however, mean that analytic solutions are possible only in special, though characteristic, moments in the lifetime of these universes. Nevertheless, our solutions exhibit persistent patterns that make us confident enough to generalise them. Thus, we confirm earlier claims that, compared to the Newtonian studies, the relativistic analysis supports considerably stronger linear growth-rates for peculiar-velocity perturbations. This result holds irrespective of the background curvature. Moreover, for positive curvature, the peculiar growth-rate is found to be faster than that obtained in a spatially flat Friedman universe. In contrast, linear peculiar velocities appear to grow at a slower pace when their Friedmann host is spatially open. Extrapolating them to the present, our results seem to suggest faster bulk peculiar motions in overdense, rather than in underdense, regions of the universe.
\end{abstract}

\section{Introduction}\label{sI}
Large-scale peculiar motions appear to be commonplace in our universe~\cite{Aetal}. The typical sizes of these bulk flows extend to few hundred Mpc and their velocities are of the order of few hundred km/sec. So far, the various surveys appear to largely agree on the direction of these motions, but not on their size and speed. There have also been reports of extreme bulk flows, which are known as ``dark flows'', with sizes and velocities well in excess of those typically anticipated (e.g.~see~\cite{KA-BKE}). Although dark flows seem at odds with the Planck data~\cite{Aetal}, a considerable number of independent reports still claim peculiar velocities in excess of those allowed by the current cosmological paradigm~\cite{WFH}-\cite{Saetal}. Among them is the very recent report of~\cite{Wetal}, which analyses the data of the \textit{CosmicFlows-4} catalogue. All these bulk flows are believed to have started as weak peculiar-velocity perturbations, initially triggered by the increasing non-uniformity of the early post-recombination era and subsequently driven by the onset of structure formation. Here, we look into the linear evolution of peculiar velocities from the theoretical viewpoint, by applying relativistic cosmological perturbation theory to ``tilted'' Friedmann-Robertson-Walker (FRW) universes (e.g.~see~\cite{TCM,EMM}). Since the tilted spacetimes allow for two (at least) groups of relatively moving observers, they are the best suited theoretical models for the relativistic study of peculiar motions in cosmological environments.\footnote{Studies involving peculiar velocities in tilted spacetimes, but without focusing on their linear evolution and growth, have also recently appeared in~\cite{NS,LB}. The former in an attempt to alleviate the Hubble tension, the latter to investigate the decay of dark matter in voids.}

When studying peculiar motions one has to distinguish the Newtonian from the relativistic treatments, even at the linear level. The reason is the fundamentally different way the two theories treat both the gravitational field and its sources. In Newtonian physics gravity is a force that is triggered and sustained by a scalar gravitational potential, where only the density of the matter contributes (via Poisson's formula). In general relativity, on the other hand, gravity is not a force but the manifestation of spacetime curvature. Moreover, it is not only the density of the matter that contributes to the gravitational field, but there is additional input from the pressure (both isotropic and viscous) and from any energy flux that may be present. Put another way, in Einstein's theory \textit{matter fluxes gravitate} as well. The flux-contribution to the energy-momentum tensor acquires special significance in studies of peculiar motions, since the latter are nothing else but matter in motion and moving matter implies nonzero energy flux.

Without accounting for the gravitational input of the peculiar flux, linear studies of peculiar-velocity ($v$) perturbations, on a homogeneous and isotropic Friedmann background, were found to grow at the rather moderate rate of $v\propto t^{1/3}$ after equipartition. These studies are primarily (purely) Newtonian in nature (e.g.~\cite{Pe}-\cite{Pa}), but they also include a few quasi-Newtonian treatments (e.g.~\cite{M,EvEM}). The latter recover the aforementioned Newtonian growth-rate, but only after imposing strict constraints upon the host spacetime, which severely compromise its relativistic nature and eventually lead to Newtonian-like equations and results (see \S~6.8.2 in~\cite{EMM} for warning comments on the use of the quasi-Newtonian approach, as well as footnote~2 here). There are also (very few) fully relativistic linear treatments of multi-component media in the literature, but they still bypass the aforementioned peculiar-flux input to the gravitational field. This happens because, after a certain stage, the analysis is switched to the so-called energy (or Landau-Lifshitz) frame, where the total flux vanishes by construction (e.g.~see \S~3.3.3 and  \S~10.4.3 in~\cite{TCM} and~\cite{EMM} respectively). In all these studies, the gravitational input of the peculiar flux is (inadvertently) unaccounted for and their results merely reproduce those of the Newtonian treatments. The interested reader is referred to~\cite{T} for a discussion on the fundamental role of the flux in studies of cosmological peculiar-velocity fields, as well as for a direct comparison between the proper relativistic treatment of the subject and the rest.

Just like any other source of energy, the flux of the peculiar motion also contributes to the linearised energy-momentum tensor of the perturbed spacetime. This in turn feeds, through the Einstein equations, into the linear energy and momentum conservation laws and eventually leads to flux-related terms in the equations of the linear cosmological perturbation theory. The related formulae follow readily from  those associated with a general imperfect fluid, the nonlinear expressions of which have been available in the literature (e.g.~see Eqs.~(1.3.21), (2.3.1) in~\cite{TCM} and/or Eqs.~(5.12), (10.101) in~\cite{EMM}). Nevertheless, the aforementioned relations were only recently utilised in cosmological studies involving peculiar velocities~\cite{TT,FT}. There, by properly accounting for the peculiar-flux input to the gravitational field, it was claimed that the relativistic analysis supports stronger linear growth-rates for peculiar-velocity perturbations. More specifically, assuming an Einstein-de Sitter background, the peculiar velocities were found to grow as $v\propto t^{4/3}$, instead of following the Newtonian/quasi-Newtonian $v\propto t^{1/3}$ law~\cite{TT,FT}. Growth, though at a slightly slower pace, was also obtained for peculiar velocities in the Cold Dark Matter (CDM) sector during the late stages of the radiation era~\cite{MT}). On the other hand, the relativistic analysis concluded that peculiar-velocity perturbations cannot survive a phase of de Sitter inflation. During that epoch all linear sources of peculiar velocities are switched off, while pre-existing velocity perturbations (if any) decay exponentially~\cite{MT}.

The present study extends those of~\cite{TT}-\cite{MT} to include FRW backgrounds with nonzero spatial curvature. In the process, we maintain the zero-pressure assumption of the previous treatments, which means that our analysis applies to times between equipartition and dark-energy domination. The technical complexities of cosmological spacetimes with non-Euclidean 3-geometry, however, allow for analytic solutions only in limiting, or special, cases. Nevertheless, there is a repeated pattern in these solutions, which suggests that (compared to the Einstein-de Sitter case) linear peculiar-velocity perturbations grow slower on open FRW backgrounds and faster on closed ones. In the former models, the growth-rate of the peculiar-velocity field reaches the Einstein-de Sitter pace ($v\propto t^{4/3}$) near the $\Omega\rightarrow1^-$ limit, as it was largely expected, but drops down to $v\propto t$ close to the Milne limit (where $\Omega\rightarrow0^+$). Although, strictly speaking, our analytic solutions cover only these two limiting cases, they do suggest that linear peculiar-velocity perturbations grow slower in open FRW universes than in their spatially flat counterparts (where $v\propto t^{4/3}$ always). Moreover, the lower the density parameter ($\Omega$) of the spatially open background, the slower the linear growth of the peculiar velocity field.

In contrast, we find that the higher density of the closed Friedmann universes can lead to peculiar growth-rates faster than the Einstein-de Sitter pace of $v\propto t^{4/3}$. The latter is recovered only near the Euclidean limit, namely when $\Omega\rightarrow1^+$. As in the open FRW models, our analytic solutions do not cover the whole lifetime of the closed universes, but apply to characteristic (though brief) moments of their post-recombination evolution. Having said that, there seems to be a clear pattern in these solutions, suggesting that the higher the density parameter of the closed universe, the stronger the linear peculiar-velocity growth. Although further study is necessary, when extrapolated to the present, these results seem to argue for faster bulk peculiar flows in overdense regions of the universe than within low-density domains.

\section{Cosmological peculiar-velocity fields}\label{sCP-VFs}
In relativity observers moving with respect to each other measure their own time and have their individual 3-dimensional rest-space. As a result, observers in relative motion experience different versions of what one might call ``reality''.

\subsection{Peculiar flux and peculiar 
4-acceleration}\label{ssPFP4A}
Cosmological peculiar velocities are defined and measured relative to the reference frame of the universe. This has been typically identified with the coordinate system of the Cosmic Microwave Background (CMB), namely with the rest-frame of the cosmic microwave photons. By definition, the latter is the only coordinate system where the radiation dipole vanishes.\footnote{In many studies, primarily (though not exclusively) of astrophysical nature, the CMB frame is also referred to as the Hubble frame. Although, strictly speaking, the two coordinate systems are distinct, it is not uncommon to find these terms being used interchangeably. Here, we will try to maintain the distinction.} This will be our assumption as well, although the generality of the following analysis implies that it still holds if peculiar velocities were to be defined relative to a different reference frame.

\begin{figure}[!tbp]
\centering\includegraphics[width=0.5\textwidth]{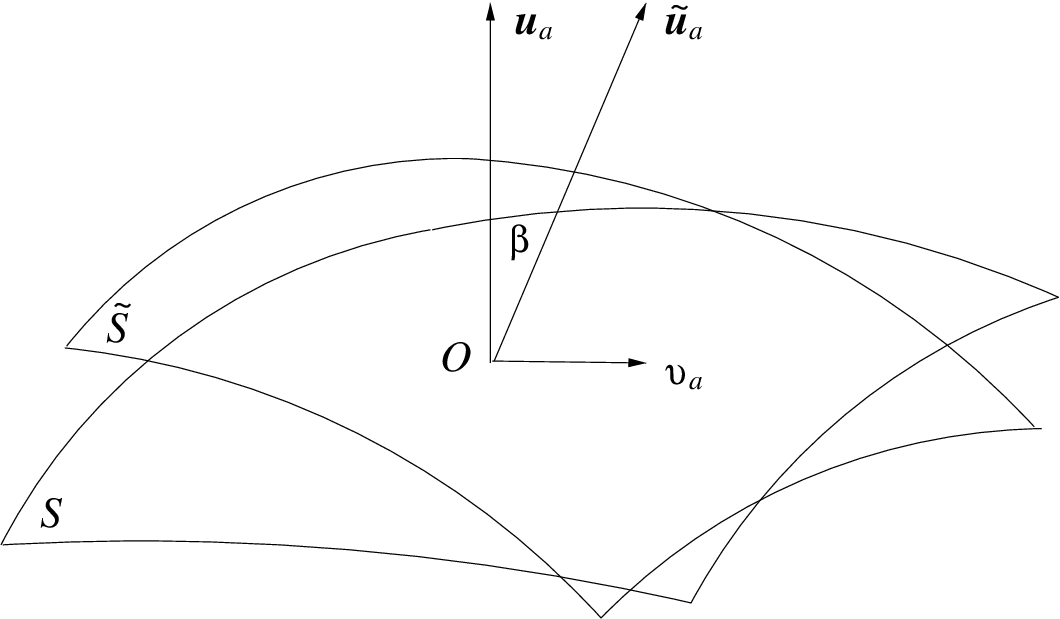}
  \caption{Tilted spacetimes allow for two families of relatively moving observers, with 4-velocities $u_a$ and $\tilde{u}_a$, at every event ($O$). Assuming that the $u_a$-field defines the reference (CMB) frame of the universe, $\tilde{u}_a$ is the 4-velocity of the matter ``drifting'' with peculiar velocity $v_a$ relative to it. Note that $\beta$ (with $\cosh\beta= -u_a\tilde{u}^a$) is the hyperbolic (tilt) angle between $u_a$ and $\tilde{u}_a$, while $S$ and $\tilde{S}$ are the 3-D rest-spaces of the two aforementioned coordinate systems.}  \label{fig:pmotion}
\end{figure}

Observers in typical galaxies do not follow the coordinate system of the microwave radiation, but the ``tilted'' frame of the matter. The latter moves with respect to its CMB counterpart with finite peculiar velocity (see Fig.~\ref{fig:pmotion}). As a result of their peculiar motion, the ``real'' observers see an apparent dipolar anisotropy in the cosmic microwave spectrum, whereas their idealised CMB counterparts see none. On these grounds, the peculiar velocity of our Local Group of galaxies has been estimated close to 600~km/sec.

Assuming that the peculiar motions are non-relativistic, the 4-velocities of the reference and the matter frames ($u_a$ and $\tilde{u}_a$ respectively) are related by $\tilde{u}_a= u_a+v_a$, with $v_a$ representing the peculiar velocity of the latter relative to the former coordinate system (see~\cite{TCM,EMM} and also Fig.~\ref{fig:pmotion} here). Note that $u_av^a=0$ always and $v^2=v_av^a\ll1$ at the non-relativistic limit.

\subsubsection{Peculiar flux}\label{sssPF}
Relative motions interfere with the nature of the cosmic medium, as experienced by the observers involved. Indeed, while the density, the isotropic pressure and the viscosity of the matter are the same in both frames at the linear level, the corresponding energy fluxes differ. In other words, $\tilde{\rho}=\rho$, $\tilde{p}=p$ and $\tilde{\pi}_{ab}=\pi_{ab}$ to first approximation, whereas~\cite{M}
\begin{equation}
\tilde{q}_a= q_a- (\rho+p)v_a\,.  \label{lrels}
\end{equation}
The above ensures that linear peculiar velocities imply nonzero \textit{peculiar flux} and vice versa. As a result, the cosmic medium behaves as an imperfect fluid, with the ``imperfection'' taking the form of an energy flux, solely due to relative-motion effects (e.g.~see \S~5.2.1 in~\cite{EMM}). Indeed, when $\tilde{q}_a=0$, we have $q_a=(\rho+p)v_a\neq0$, while $q_a=0$ leads to $\tilde{q}_a=-(\rho+p)v_a\neq0$.\footnote{Hereafter, tildas will always denote variables and operators evaluated in the ``tilted'' frame of the matter.} Clearly, setting both fluxes to zero in Eq.~(\ref{lrels}), makes the peculiar-velocity field disappear. The only exception is on FRW backgrounds with a de Sitter inflationary ($p=-\rho$) equation of state. There, linear velocity perturbations either vanish identically, or decay exponentially (see~\cite{MT} for further discussion and technical details).

In relativity, contrary to Newtonian physics, the energy flux contributes to the energy-momentum tensor and therefore to the gravitational field.  As a result, when peculiar motions are present, there is a nonzero linear 4-acceleration, even in the absence of pressure.

\subsubsection{Peculiar 4-acceleration}\label{sssP4A}
Nonzero peculiar flux means that the cosmic medium can no longer be treated as perfect (see above and also~\cite{EMM}). Imperfect fluids have nonzero linear 4-acceleration, even when their pressure is zero (e.g.~see below and also~\cite{TCM,EMM}). Put another way, a nonzero \textit{peculiar flux} implies a nonzero \textit{peculiar 4-acceleration}, even in the absence of pressure. The direct interconnection between these two physical entities follows from the aforementioned gravitational input of the peculiar flux, which feeds into the Einstein equations and then emerges in the relativistic energy and momentum conservation laws. Indeed, assuming zero pressure and an FRW background, the two conservation laws linearise to
\begin{equation}
\dot{\rho}= -3H\rho- {\rm D}^aq_a \hspace{10mm} {\rm and} \hspace{10mm} \rho A_a= -\dot{q}_a-4Hq_a\,,  \label{lcls}
\end{equation}
respectively. Here, $A_a$ is the 4-acceleration measured in the reference (CMB) coordinate system and $H$ is the Hubble parameter of the Friedmann background. Also, overdots indicate time derivatives and ${\rm D}_a$ is the spatial (covariant) derivative operator in the $u_a$-frame. Note that, in deriving the above, we have assumed that there is no flux in the coordinate system of the matter, which (in the absence of pressure) moves along timelike geodesics. Technically speaking, we have adopted the conventions of~\cite{M,EvEM}, by setting $\tilde{q}_a=0=\tilde{A}_a$ in the matter frame.

Although the momentum conservation law (\ref{lcls}b) guarantees the presence of the 4-acceleration, the form of $A_a$ follows from relativistic cosmological perturbation theory. More specifically, taking the spatial gradient of the energy-density conservation equation (\ref{lcls}a) and employing the linear commutation law ${\rm D}_a\dot{\rho}=({\rm D}_a\rho)^{\cdot}+H{\rm D}_a\rho+3H\rho A_a$, one arrives at
\begin{equation}
A_a= {1\over3H}\,{\rm D}_a\vartheta- {1\over3aH}\left(\dot{\Delta}_a+\mathcal{Z}_a\right)\,,  \label{lA}
\end{equation}
Here, $\vartheta={\rm D}^av_a$ is the divergence of the peculiar-velocity field, while $\Delta_a$ and $\mathcal{Z}_a$ monitor inhomogeneities in the matter distribution and in the expansion respectively~\cite{TCM,EMM}. This is the fully relativistic linear expression for the 4-acceleration in the presence of peculiar velocities.

The gravitational input of the peculiar flux and the resulting peculiar 4-acceleration are purely relativistic effects that ``survive'' at the linear perturbative level, but they are both bypassed (for different reasons) in Newtonian and quasi-Newtonian, treatments.\footnote{In Newtonian theory only the density of the matter contributes to gravity, which means that the peculiar flux (although nonzero) has no gravitational input by default. Then, in the absence of pressure, the Newtonian acceleration is given by the gradient of the gravitational potential. The quasi-Newtonian studies have an apparently relativistic profile, which however is severely compromised by their constraints (see \S~6.8.2 in~\cite{EMM} for related discussion and ``warning'' comments). In particular, by adopting a reference frame with zero linear shear and vorticity, the perturbed quasi-Newtonian spacetime has no gravitational waves either. Crucially, the 4-acceleration is not given by the relativistic expression (\ref{lA}), but by the gradient $A_a={\rm D}_a\varphi$, where $\varphi$ is a scalar ``gravitational'' potential essentially identical to its Newtonian analogue~\cite{M,EvEM}. All this may simplify the mathematics, but it compromises the physics, since the relativistic gravitational input of the peculiar flux is not accounted for. It should therefore come to no surprise that both the Newtonian and the quasi-Newtonian studies of cosmological peculiar velocities arrive at the same linear growth of $v\propto t^{1/3}$. Here, as well as in the relativistic studies of~\cite{TT,MT}, no restrictions are imposed and our reference frame has nonzero shear and vorticity at the linear level. Nevertheless, none of these variables is involved in the linear calculations, which makes them irrelevant for the purposes of this study. What makes the difference is the gravitational input of the peculiar flux, which means that the linear 4-acceleration is given by Eq.~(\ref{lA}), instead of the gradient of an arbitrary scaler potential. We refer the reader to~\cite{T} for the comparison between the proper relativistic study of peculiar motions and the rest.} Accounting for the peculiar-flux and for the peculiar 4-acceleration is what separates the relativistic studies of peculiar motions from the rest.

\section{Peculiar velocities in FRW universes}\label{sPVFRWUs}
The observed large-scale bulk flows are believed to have started as weak peculiar velocity perturbations that were driven and enhanced  by structure formation, namely by the ever increasing inhomogeneity and anisotropy of the post-recombination universe.

\subsection{Relativistic linear evolution}\label{ssRLE}
The role of structure formation in driving peculiar-velocity perturbations is reflected in their evolution formula. The latter can be obtained from the momentum conservation law (\ref{lcls}b), which recasts into the linear form
\begin{equation}
\dot{v}_a+ Hv_a= -A_a\,,  \label{ldotv}
\end{equation}
after taking into account that $q_a=\rho v_a$ in our case (see Eq.~(\ref{lrels}) and recall that we have set $p=0=\tilde{q}_a$). Therefore, linear peculiar-velocity perturbations are driven by the 4-acceleration. The acceleration/4-acceleration is the driving force of linear peculiar velocities in the Newtonian/quasi-Newtonian studies as well, although there it is given by the gradient of a scalar potential and not by the fully relativistic expression (\ref{lA}). The latter also ensures that the actual driving agents are non-gravitational forces (represented by $\Delta_a$ and $\mathcal{Z}_a$) triggered by structure formation.

Differentiating (\ref{lA}) and (\ref{ldotv}) in time, combining the resulting expressions and using the linear commutation law $({\rm D}_a\vartheta)^{\cdot}={\rm D}_a \dot{\vartheta}+H{\rm D}_a\vartheta$, leads to the differential equation~\cite{TT}-\cite{MT}
\begin{equation}
\ddot{v}_a+ \left(1-{1\over2}\,\Omega\right)H\dot{v}_a- H^2(1+\Omega)v_a= {1\over3H}\,{\rm D}_a\dot{\vartheta}+ {1\over3aH} \left(\ddot{\Delta}_a+\dot{\mathcal{Z}}_a\right)\,,  \label{lddotv1}
\end{equation}
given that $\dot{H}=-H^2(1+\Omega/2)$ in the FRW background (with $\Omega=\kappa\rho/3H^2$ being the associated density parameter). The above monitors the linear evolution of peculiar velocities in the pressureless matter within a tilted almost-FRW universe.  One may therefore apply (\ref{lddotv1}) to baryonic ``dust'' after recombination and to low-energy CDM species after, as well as before, decoupling.

\subsection{The case of a flat FRW universe}\label{ssCFFRWU}
The inhomogeneous differential equation (\ref{lddotv1}) has no analytic solution, even when the Friedmann background is spatially flat. By confining to the homogeneous component of (\ref{lddotv1}) and assuming an Einstein-de Sitter background (where $\Omega=1$ and $H=2/3t$), the latter reduces to.\footnote{The inhomogeneity/homogeneity of (\ref{lddotv1}) refers to the nature of the specific differential equation and has nothing to do with the symmetries of the host spacetime.}
\begin{equation}
\ddot{v}_a+ {1\over3t}\,\dot{v}_a- {8\over9t^2}\,v_a= 0\,,  \label{lddotv2}
\end{equation}
giving
\begin{equation}
v= \mathcal{C}_1t^{4/3}+ \mathcal{C}_2t^{-2/3}= \mathcal{C}_3a^2+ \mathcal{C}_4a^{-1}\,,  \label{lv1}
\end{equation}
since $a\propto t^{2/3}$ after equipartition~\cite{TT}-\cite{MT}. The growth rate of the peculiar-velocity field (i.e.~$v\propto t^{4/3}\propto a^{2}$) obtained here is considerably stronger than that of the Newtonian and quasi-Newtonian treatments, where $v\propto t^{1/3}\propto a^{1/2}$.\footnote{Theory tells us that the solution of an inhomogeneous differential equation forms by the full homogeneous solution plus a partial inhomogeneous solution. Therefore, solving (\ref{lddotv1}) in full, can make practical difference only when the partial solution grows stronger than the strongest growing mode of its homogeneous counterpart.}

\section{Peculiar velocities in curved FRW
universes}\label{ssPVCFRWUs}
The geometry of the Friedmann background can in principle change the evolution of the peculiar velocity field. Such changes may also apply to locally overdense, or/and underdense, regions of perturbed FRW universes, which may still lie close to the Euclidean limit globally.

\subsection{The linear equations}\label{ssLEs}
When the Friedmann models have nonzero spatial curvature, one needs to recast the evolution equations in terms of the conformal time ($\eta$, with $\dot{\eta}=1/a$) instead of proper time (e.g.~see~\cite{N} for further discussion). This is achieved by means of the transformation formulae
\begin{equation}
\dot{v}= {1\over a}{{\rm d}v\over{\rm d}\eta} \hspace{15mm} {\rm and} \hspace{15mm} \ddot{v}= {1\over a^2}\left({{\rm d}^2v\over{\rm d}\eta^2}- \mathcal{H}{{\rm d}v\over{\rm d}\eta}\right)\,,  \label{tls}
\end{equation}
with $\mathcal{H}=aH$ representing the ``rescaled'' Hubble parameter. Employing the above, the homogeneous component of Eq.~(\ref{lddotv1}) recasts as
\begin{equation}
{{\rm d}^2v\over{\rm d}\eta^2}- {1\over2}\,\Omega\mathcal{H} {{\rm d}v\over{\rm d}\eta}- (1+\Omega)\mathcal{H}^2v= 0\,.  \label{K1ddotv1}
\end{equation}

Assuming a spatially open Friedmann background filled with pressureless dust (baryonic or not), we have $a\propto\sinh^2(\eta/2)$ with $\eta>0$. Then, $\mathcal{H}=\coth(\eta/2)$ and the differential equation (\ref{K1ddotv1}) takes the form
\begin{equation}
{{\rm d}^2v\over{\rm d}\eta^2}- {1\over2}\,\Omega\coth
\left({\eta\over2}\right){{\rm d}v\over{\rm d}\eta}- (1+\Omega)\coth^2\left({\eta\over2}\right)v= 0\,,  \label{-1ddotv1}
\end{equation}
where $0<\Omega<1$ always. Alternatively, when the unperturbed FRW spacetime has closed spatial sections, the scale factor is given by  $a\propto\sin^2(\eta/2)$ (with $0<\eta<2\pi$), while the rescaled Hubble parameter is $\mathcal{H}=\cot(\eta/2)$. In this environment, expression (\ref{K1ddotv1}) reads
\begin{equation}
{{\rm d}^2v\over{\rm d}\eta^2}- {1\over2}\,\Omega\cot\left({\eta\over2}\right){{\rm d}v\over{\rm d}\eta}- (1+\Omega)\cot^2\left({\eta\over2}\right)v= 0\,,  \label{+1ddotv1}
\end{equation}
with $\Omega>1$ at all times. Note that the sign of the second term on the left-hand side of the above changes when the closed FRW background enters its contracting phase. More specifically, the coefficient of the aforementioned term is negative during expansion (i.e.~when $0<\eta<\pi$) and positive when the universe is collapsing (i.e.~for $\pi<\eta<2\pi$). This suggests that linear peculiar velocities evolve differently during these two phases (see more in \S~\ref{ssCCFRWU} below).

\subsection{Close to the Euclidean limit}\label{ssCEL}
The differential equations (\ref{-1ddotv1}) and (\ref{+1ddotv1}) monitor the linear evolution of peculiar-velocity perturbations in open and closed almost-FRW universes respectively. Before we proceed to obtain solutions, it is worth verifying that solving both of the above at the Euclidean ($\Omega\rightarrow1^{\mp}$) limit recovers solution (\ref{lv1}) obtained in~\cite{TT}-\cite{MT} and reproduced in \S~\ref{ssCFFRWU} here. Indeed, marginally open/closed Friedmann universes containing conventional matter (i.e.~with $\rho+3p>0$) correspond to the $\eta\rightarrow0^+$ limit.\footnote{In FRW universes with nonzero spatial curvature, the Friedmann equation takes the form $|1-\Omega|=|K|/\mathcal{H}$, where $K=\pm1$ is the 3-curvature index. Evaluated in an open model after equipartition, when $\mathcal{H}=\coth(\eta/2)$, the latter recasts into $|1-\Omega|=\tanh^2(\eta/2)$. Consequently, the $\Omega\rightarrow1^-$ limit corresponds to $\eta\rightarrow0^+$. The same also holds in spatially closed FRW cosmologies, where $|1-\Omega|=\tan(\eta/2)$ and therefore $\eta\rightarrow0^+$ when $\Omega\rightarrow1^+$.} Then, a simple Taylor-series expansion gives $\coth(\eta/2)=2/\eta=\cot(\eta/2)$ to first approximation, all of which recast Eqs.~(\ref{-1ddotv1}) and (\ref{+1ddotv1}) into the single differential formula
\begin{equation}
{{\rm d}^2v\over{\rm d}\eta^2}-{1\over\eta}{{\rm d}v\over{\rm d}\eta}-{8\over\eta^2}\,v= 0\,.  \label{mK1ddotv1}
\end{equation}
The latter accepts the power-law solution
\begin{equation}
\tilde{v}= \mathcal{C}_1\eta^4+ \mathcal{C}_2\eta^{-2}\,,  \label{mKv1}
\end{equation}
expressed in terms of the conformal time. Recalling that $t\propto\eta-\sinh(\eta)$ and $t\propto\eta-\sin(\eta)$ in open and closed FRW cosmologies respectively (e.g.~see~\cite{N}), a Taylor-series expansion at the $\eta\rightarrow0^+$ limit gives $t\propto\eta^3$ to leading order in both cases. Then, solution (\ref{mKv1}) transforms into
\begin{equation}
v= \mathcal{C}_3t^{4/3}+ \mathcal{C}_4t^{-2/3}= \mathcal{C}_5a^2+ \mathcal{C}_6a^{-1}\,,  \label{mKv2}
\end{equation}
in agreement with (\ref{lv1}) and with~\cite{TT}-\cite{MT}. Note that the scale-factor dependence seen in the second equality follows from the fact that $a\propto1-\cosh(\eta)$ and $a\propto1-\cos(\eta)$ in open and closed Friedmann universes respectively~\cite{N}. Then, at the $\eta\rightarrow0^+$ limit, we obtain $a\propto\eta^2\propto t^{2/3}$ to first approximation in both marginally open and marginally closed models (as expected).

\subsection{The case of an open FRW universe}\label{ssCOFRWU}
Friedmann universes with hyperbolic spatial geometry are infinite in time, with $\eta\in(0,+\infty)$. The open FRW spacetimes are infinite in space as well, unless their topology is nontrivial. Next we will focus on the late-time evolution of these models.

\subsubsection{Late-time solutions}\label{sssL-TSs}
When dealing with spatially open FRW universes, it is generally impossible to obtain analytic solutions that cover the whole evolution of these models. Analytic solutions are usually possible at late times, namely when $\eta\gg1$. At this limit, $\coth(\eta/2)\rightarrow1$ and Eq.~(\ref{-1ddotv1}) simplifies to
\begin{equation}
{{\rm d}^2v\over{\rm d}\eta^2}- {1\over2}\,\Omega{{\rm d}v\over{\rm d}\eta}- (1+\Omega)v= 0\,.  \label{-1ltlddottv2}
\end{equation}
Assuming that $\Omega$ is (nearly) constant during the late evolutionary period under consideration, the above solves analytically to give
\begin{equation}
v= \mathcal{C}_1{\rm e}^{\alpha_{_1}\eta}+  \mathcal{C}_2{\rm e}^{\alpha_{_2}\eta}\,,  \label{-1ltlv1}
\end{equation}
with
\begin{equation}
\alpha_{1,2}= {1\over4}\left(\Omega\pm\sqrt{\Omega^2+16\Omega+16}\right)  \label{alphas1}
\end{equation}
and $0<\Omega<1$. Consequently, $\alpha_1>0$ and $\alpha_2<0$ to guarantee that solution (\ref{-1ltlv1}) contains one growing and one decaying mode always. Moreover, recalling that $t\propto\sinh(\eta)-\eta$ in open FRW cosmologies and that $\sinh(\eta)=({\rm e}^{\eta}-{\rm e}^{-\eta})/2$ by definition, we deduce that $t\propto{\rm e}^{\eta}$ at late times (i.e.~when $\eta\gg1$). Then, expressed in terms of proper time, solution (\ref{-1ltlv1}) reads
\begin{equation}
v= \mathcal{C}_3t^{\alpha_{_1}}+  \mathcal{C}_4t^{\alpha_{_2}}\,,  \label{-1ltltv2}
\end{equation}
where $\alpha_{1,2}$ are still given by (\ref{alphas1}). The growing mode of the latter (corresponding to $\alpha_1= (\Omega+\sqrt{\Omega^2+16\Omega+16})/4$) ensures that the lower the background density parameter, the slower the linear growth of peculiar-velocity perturbations. Nevertheless, as long as $\Omega>0$, expression (\ref{alphas1}) also guarantees that $\alpha_1>1$ to ensure that the peculiar-velocity field grows faster than $v\propto t^{1/3}$, which is the growth rate obtained by the Newtonian and the quasi-Newtonian studies. On the other hand, it is straightforward to show that $\alpha_1<4/3$ as long as $\Omega<1$. Recall that $v\propto t^{4/3}$ is the linear growth-rate of peculiar-velocity perturbations on flat FRW backgrounds, obtained by the relativistic analysis (see~\S~\ref{ssCFFRWU} here and also~\cite{TT}-\cite{MT}).

\subsubsection{Close to the Milne limit}\label{sssCML}
Before closing this section, it worth noticing that at the very late stages of its expansion an open FRW universe approaches the empty Milne solution. Then, taking the $\Omega\rightarrow0^+$ limit of (\ref{alphas1}), we obtain $\alpha_{1,2}=\pm1$ and consequently the power-law evolution
\begin{equation}
v= C_3t+ C_4t^{-1}\,,  \label{Milne}
\end{equation}
for the peculiar-velocity field. Not surprisingly, the same result also holds when the background model is a Milne universe at all times. Indeed, setting $\Omega=0$ in Eq.~(\ref{lddotv1}) and keeping in mind that $a=t$ in the Milne cosmology, one immediately recovers the above solution.\footnote{In the absence of matter, it may sound rather unphysical to talk about peculiar velocities. In this respect, our Milne limit corresponds to a nearly empty (perturbed) FRW background (with $\Omega=0$ to zero order only).}

Overall, the (relativistic) linear growth-rates of peculiar-velocity perturbations  on open FRW backgrounds are faster than that of the Newtonian/quasi-Newtonian analysis (where $v\propto t^{1/3}$), but do not exceed the rate obtained by applying relativistic perturbation theory to an Einstein-de Sitter background (where $v\propto t^{4/3}$ -- see solution (\ref{lv1}) in \S~\ref{ssCFFRWU} above). The latter is reached only at the $\Omega\rightarrow1^-$ limit. In fact, the lower the density parameter of the open background, the weaker the linear growth-rate of the $v$-field was found to be. Strictly speaking, however, and in the absence of an analytic solution of Eq.~(\ref{-1ddotv1}) for the whole lifetime of the open FRW universes, the aforementioned conclusions apply to their late-time stages only. Having said that, it sounds plausible that the same $\Omega$-dependence of the linear peculiar-velocity growth-rate should hold at all times on Friedmann backgrounds with less than the critical density.

\subsection{The case of a closed FRW universe}\label{ssCCFRWU}
Spatially closed Friedmann models are finite both in space and in time (conformal as well as proper). More specifically, when matter is pressure-free dust, the conformal time is bounded within the interval $\eta\in[0,2\pi]$. This means that we cannot take the $\eta\gg1$ limit of Eq.~(\ref{+1ddotv1}), as we did in the case of the open model before. We are therefore ``forced'' to look for analytic solutions at characteristic moments in the lifetime of a closed FRW universe.

\subsubsection{Close to the Einstein-static limit}\label{sssCE-SL}
Before doing so, let us first consider a special case of the closed Friedman spacetimes, namely the Einstein-static universe. The latter has long been known to be unstable to linear spatially homogeneous and isotropic density perturbations~\cite{E}, though the stability issue is less straightforward when dealing with conformal metric and inhomogeneous distortions~\cite{H}-\cite{BEMT}. Allowing for linear peculiar-velocity perturbations on an Einstein-static background (with $H=0$ to zero order), the homogeneous component of differential equation (\ref{lddotv1}) reduces to $\ddot{v}_a=0$. The later solves immediately to give $v\propto t\propto\eta$ on all scales, confirming that the Einstein universe is unstable to linear peculiar-velocity perturbations. Recall also that $t\propto\eta$ to first approximation on an Einstein-static background.

In an analogous fashion, one can consider perturbations around characteristic brief ``moments'' in the evolution of a closed FRW universe. Here, these brief moments will correspond to the values $\eta=\pi/2$, $\eta=3\pi/2$  and $\eta=\pi$ of the conformal time. The first of the above marks the middle of the expanding phase, the second applies half the way down the subsequent contracting epoch and the third corresponds to the ``turning point'' of maximum expansion. Consequently, the resulting solutions describe the linear evolution of peculiar-velocity perturbations in the ``vicinity'' of these characteristic states.

\subsubsection{Expanding-phase solutions}\label{sssE-PSs}
Focusing on the expanding phase and setting $\eta=\pi/2$, gives $\cot(\eta/2)=1$ at the mid-expansion point. This in turn reduces Eq.~(\ref{+1ddotv1}) to
\begin{equation}
{{\rm d}^2v\over{\rm d}\eta^2}- {1\over2}\,\Omega{{\rm d}v\over{\rm d}\eta}- (1+\Omega)v= 0\,, \label{+1ddotv2}
\end{equation}
with $\Omega>1$. Since we are considering a brief period near the middle of the expansion phase, we may treat $\Omega$ as almost constant. Then, the above accepts the power-law solution
\begin{equation}
v= \mathcal{C}_1{\rm e}^{\alpha_{_1}\eta}+  \mathcal{C}_2{\rm e}^{\alpha_{_2}\eta}\,,  \label{+1ltv1}
\end{equation}
where
\begin{equation}
\alpha_{1,2}= {1\over4}\left(\Omega\pm\sqrt{\Omega^2+16\Omega+16}\right)\,,  \label{alphas2}
\end{equation}
to guarantee again one growing and one decaying mode (with $\alpha_1>0$ and $\alpha_2<0$ respectively). The growing mode also ensures that the higher the value of the density parameter, the stronger the linear growth of the peculiar-velocity field. Note that (\ref{+1ltv1}) is formally identical to the late-time solution obtained in spatially open FRW universes (compare to Eqs.~(\ref{-1ltlv1}), (\ref{alphas1}) in \S~\ref{ssCOFRWU}). Nevertheless, there is a key difference between the two results, because $\Omega<1$ in the case of an open universe and $\Omega>1$ here. This allows for growth rates stronger than those obtained both in flat and in open FRW models. In particular, when $\Omega\gg1$, solution (\ref{+1ltv1}), (\ref{alphas2}) ensures that $\alpha_2\simeq(\Omega+4)/2\gg1$. The latter suggests an arbitrarily strong linear growth for the peculiar-velocity field, when the background density parameter increases arbitrarily as well.

\subsubsection{Contracting-phase solutions}\label{sssC-PSs}
The half-way point down the contracting phase of a spatially closed Friedmann model corresponds to $\eta=3\pi/2$. There, Eq.~(\ref{+1ddotv1}) reads
\begin{equation}
{{\rm d}^2v\over{\rm d}\eta^2}+ {1\over2}\,\Omega{{\rm d}v\over{\rm d}\eta}- (1+\Omega)v= 0\,, \label{+1tv''3}
\end{equation}
giving
\begin{equation}
v= \mathcal{C}_1{\rm e}^{\alpha_{_1}\eta}+  \mathcal{C}_2{\rm e}^{\alpha_{_2}\eta}\,,  \label{+1ltv2}
\end{equation}
where now
\begin{equation}
\alpha_{1,2}= {1\over4}\left(-\Omega\pm\sqrt{\Omega^2+16\Omega+16}\right)\,.  \label{alphas3}
\end{equation}
As before, $\alpha_1$ is positive and corresponds to the growing mode, whereas $\alpha_2$ is negative and denotes its decaying counterpart. Following (\ref{+1ltv1}) and (\ref{+1ltv2}), the larger/smaller the value of $\Omega$, the stronger/weaker the linear growth of peculiar-velocity perturbations.

Comparing solutions (\ref{+1ltv1}), (\ref{alphas2}) and (\ref{+1ltv2}), (\ref{alphas3}), we notice that the growth-rate of the peculiar-velocity perturbations is stronger in the expanding phase than in the contracting. This reflects the fact that in closed Friedmann models the density parameter increases only during their expansion and decreases when they contract (recall that $\dot{\Omega}=-H(1-\Omega)\Omega$ for dust). As a result, unlike solution (\ref{+1ltv1})-(\ref{alphas2}), the growing mode of  (\ref{+1ltv2})-(\ref{alphas3}) cannot increase arbitrarily with $\Omega$, but has a finite upper bound. Indeed, assuming a highly dense FRW background, with $\Omega\gg1$ near the middle of its contraction, we find that $\alpha_1\lesssim2$.

\subsubsection{Close to the maximum expansion point}\label{sssCMEP}
Finally, at the moment of maximum expansion (corresponding to $\eta=\pi$), we have $\cot(\eta/2)=0$ (and $\mathcal{H}=0$ as expected). In the vicinity of that ``turning'' point, Eq.~(\ref{+1ddotv1}) gives $\tilde{v}=\tilde{C}_1+\tilde{C}_2\eta$, implying linear growth (in terms of conformal time) for the peculiar-velocity field. It comes to no surprise that this solution, which by the way is independent of the value of $\Omega$ at the time, also holds when the closed FRW background is replaced by the static Einstein cosmology (see discussion in \S~\ref{sssCE-SL} earlier).

In summary, linear peculiar-velocity perturbations can grow faster in closed FRW universes than in their flat counterparts and even faster than in the spatially open ones. Strictly speaking, this is still a tentative claim, since the analytic solutions presented here do not cover the whole lifetime of Friedmann models with $\Omega\lessgtr1$. Nevertheless, there is a persistent pattern in all of our solutions, which supports this claim. Moreover, even if the universe is close to the Euclidean limit globally, our results suggest that peculiar velocities should grow faster and stronger in locally overdense regions than in underdense local domains.

\section{Discussion}\label{sD}
The large-scale bulk peculiar motions observed in the universe today are believed to have started as weak peculiar-velocity perturbations soon after the beginning of the dust-dominated epoch of the universe. Driven by the ongoing process of structure formation, these weak velocity fields have grown to speeds of a few hundred km/sec today and have come to encompass regions of few hundred Mpc. Currently, the various surveys seem to agree on the direction of the observed bulk flows, which is more-or-less in line with that of the CMB dipole, but they appear to disagree on their speed and size. Although several surveys report bulk peculiar motions within the broad limits set by the current cosmological paradigm, there has been an increasing number of claims for peculiar flows faster than those typically allowed by the $\Lambda$CDM model. Not to mention the so-called dark flows, which are well in excess both is speed and in size. With this in mind, one wonders whether there could be a theoretical explanation to the bulk-flow question.

Following the work of~\cite{TT}-\cite{MT}, the answer could lie in the gravitational theory used to study the evolution of peculiar velocities in cosmology. More specifically, employing relativistic cosmological perturbation theory, the linear growth-rate of the peculiar-velocity field was found to be considerably stronger than that predicted by the Newtonian studies. The reason was the fundamentally different way the two theories treat the gravitational field and its sources. Whereas only the density of the matter gravitates in Newtonian physics, in general relativity energy fluxes gravitate as well. This difference plays a pivotal role when studying peculiar flows, since the latter are nothing else but matter in motion.

By default, there is no gravitational input from the peculiar flux in the Newtonian studies, which have led to a rather slow ($v\propto t^{1/3}$) growth-rate for the linear peculiar-velocity field~\cite{Pe}-\cite{Pa}. The same result was also obtained by the quasi-Newtonian treatments~\cite{M,EvEM}. However, although the latter approaches may have the external appearance of a relativistic study, they reduce to Newtonian because of the severe restrictions imposed on the (perturbed) host spacetime (see footnote~3 here and references therein). Among others, the relativistic contribution of the peculiar flux to the gravitational field is inadvertently bypassed and the quasi-Newtonian linear solutions are identical to those of the purely Newtonian analysis. In a proper relativistic study, on the other hand, the peculiar flux contributes to the linear energy-momentum tensor and, in so doing, modifies the equations monitoring the linear evolution of peculiar velocities. The result was a growth proportional to $v\propto t^{4/3}$, instead of the Newtonian/quasi-Newtonian $v\propto t^{1/3}$ rate (see~\cite{T} for a direct comparison between the relativistic and the Newtonian/quasi-Newtonian analysis of peculiar velocities). Similar, though slightly slower, growth rates have also been claimed for linear peculiar-velocity perturbations in the CDM component during the late stages of the radiation era~\cite{MT}.

Here we have revisited the bulk-flow question, by extending the relativistic study to include FRW models with nonzero spatial curvature. Maintaining the zero-pressure constraint of the previous treatments, our aim was to investigate whether and to what extent the lower/highher density of the open/closed Friedmann backgrounds changes the aforementioned flat-FRW picture. Introducing conformal time, we derived the linear formulae governing the peculiar-velocity field on open and closed FRW backgrounds. The technical complexity of these models, however, meant that analytic solutions were possible only in limiting and/or in special cases. Nevertheless, the solutions have revealed a persistent pattern that provided some physical intuition about the way peculiar-velocity perturbations behave in underdense/overdense environments.

According to our results, linear peculiar velocities grow slower on underdense FRW backgrounds with negative spatial curvature, than on their flat counterparts. More specifically, the linear velocity growth-rate is approaching that of the Einstein-de Sitter universe (i.e.~$v\propto t^{4/3}$) near the Euclidean limit, namely as $\Omega\rightarrow1^-$. At the other end, that is close to the Milne limit (where $\Omega\rightarrow0^+$ and the negative curvature dominates), the $v$-field was found to grow linearly with time (i.e.~$v\propto t$). In general, our solutions suggested that the lower the density parameter of the open universe, the weaker the linear peculiar growth-rate. Nevertheless, the latter was always found to exceed the $v\propto t^{1/3}$ rate of the Newtonian/quasi-Newtonian studies.

Turning to spatially closed Friedmann backgrounds, we found that they lead to linear growth rates stronger than those of their flat counterparts and therefore even stronger than those obtained in perturbed open FRW models. Although analytic solutions were feasible only at special (though characteristic) moments in the lifetime of these closed backgrounds, there was a persistent pattern suggesting than the higher the density parameter, the stronger the peculiar growth. An additional decisive factor was whether the universe was in its expanding or in its contracting phase. A key difference distinguishing these two periods in the lifetime of a closed FRW model is that in the former phase the density parameter increases, but it drops once the contraction starts. The direct consequence of this change in the evolution of $\Omega$ was that, while the peculiar-velocity field could grow arbitrarily strong during the expansion period, it remained bounded in the subsequent phase of universal contraction.

Overall and  despite the fact that our results do not cover the whole lifetime of the background models, a repeated pattern emerges in our linear solutions. The latter suggest that peculiar velocities grow stronger in the overdense environments of the closed Friedmann universes, rather than in those of their low-density open counterparts. On these grounds, one may expect to ``see'' faster bulk peculiar flows in high-density regions of the universe, than in underdense ``voids''. Also important is that, in all cases, the relativistic treatment has led to linear peculiar growth-rates stronger than those obtained by the Newtonian/quasi-Newtonian studies. This result provides theoretical support to an increasing number of surveys reporting large-scale peculiar motions faster that it is generally anticipated.\\

\textbf{Acknowledgements:} CGT acknowledges support from the Hellenic Foundation for Research and Innovation (H.F.R.I.), under the ``First Call for H.F.R.I. Research Projects to support Faculty members and Researchers and the procurement of high-cost research equipment Grant'' (Project Number: 789).\\

\end{document}